\documentstyle[aps,prl,epsf,floats]{revtex}
\draft

\newcommand{\beq}{\begin{equation}}
\newcommand{\eeq}{\end{equation}}
\newcommand{\beqs}{\begin{eqnarray}}
\newcommand{\eeqs}{\end{eqnarray}}

\begin{document}

\twocolumn[\hsize\textwidth\columnwidth\hsize\csname
@twocolumnfalse\endcsname]

\title{Comment on ``Observability of the Neutrino Charge Radius''}

\author{K. Fujikawa $^1$ and R. Shrock$^2$} \address{$^1$ Department of
Physics, University of Tokyo, Bunkyo-ku,Tokyo 113,Japan \\ $^2$ Yang Institute
for Theoretical Physics, State University of New York, Stony Brook, NY 11790
USA}

\maketitle

Electromagnetic properties of neutrinos are of fundamental importance and
serve as a probe of physics beyond the Standard Model (SM).  The conventional
understanding of the neutrino charge radius (NCR) is that it is not a physical
quantity \cite{bgl,ls}, as shown by the fact that it is gauge-dependent
\cite{ls}.  In \cite{bpv} (denoted BPV; see also \cite{bpv2}), BPV claim that
they can extract a gauge-independent NCR which is, therefore, a physical
observable.  We show here that the BPV claim is incorrect by demonstrating that
the NCR is, in general, gauge-dependent.  To examine the BPV claim, we use
their simplification of neglecting neutrino masses (so that Dirac and Majorana
neutrinos are equivalent and there is no lepton mixing).  The relevant matrix
element is then $\langle \nu_{\ell}
(p^\prime)|J_{em,\alpha}|\nu_{\ell}(p)\rangle = \bar \nu_{\ell}(p^\prime)
\gamma_\alpha F_1(t)(1-\gamma_5) \nu_{\ell}(p)$, and the NCR is $\langle
r^2_{\nu_\ell}\rangle = 6 dF_1(t)/dt|_{t=0} \equiv 6F_1^\prime(0)$, where
$q=p-p^\prime$ and $t=q^2$. BPV consider the amplitude $A=A_{tree} +
A_{1-loop}$ for the reaction $\nu_{\mu}(p) + e_R(k) \to \nu_{\mu}(p^{\prime}) +
e_R(k^{\prime})$ in the high-energy limit where the electron mass can be
neglected on external lines so that $+$ helicity $= R$ chirality, where
$\psi_{L,R} \equiv \frac{1}{2}(1 \mp \gamma_5)\psi$.  In the SM the 2-$W$
exchange diagrams then vanish.  Now $d\sigma/dt = |A_{tree}|^2 + 2
Re(A_{tree}A_{1-loop}^*)$ plus higher-order terms.  To isolate terms, BPV
consider the sum \cite{ssm} $d\sigma(\nu_\mu e_R \to \nu_\mu e_R)/dt +
d\sigma(\bar\nu_\mu e_R \to \bar\nu_\mu e_R)/dt$; equivalently we consider
$d\sigma(\nu_\mu e_R \to \nu_\mu e_R)/dt + d\sigma(\nu_\mu e^c_L \to \nu_\mu
e^c_L)/dt$ (specifically the $t \to 0$ limit). Since for the second term,
$A_{tree}$ reverses sign but the 2-$Z$ exchange graphs do not, this sum removes
terms from the 2-$Z$ graphs (which are gauge-invariant by themselves). So for
this sum, the $A_{1-loop}$ terms arise from the graphs involving $\gamma$
exchange with $\nu_\ell$ vertex correction (VC) (we give graphs in $U$ gauge
but have done the analysis in $R_\xi$ gauge \cite{fls}) and $Z$-exchange with
(i) $\nu_\mu$ VC, (ii) $e_R$ VC, and (iii) $Z$ propagator correction, and a
graph with a $\gamma$ coupling to $e_R$, mixing to form a $Z$ which couples to
$\nu_\mu$.  In this way, one can extract the gauge-invariant quantity
\beqs
& & A_{1-loop}=e^2 \biggl [ \frac{ F_1(t)}{t}
+ \frac{F^{\nu\nu}_Z(t) + F^{ee}_Z(t)}{t-M_Z^2}
- \frac{\Pi_{ZZ}(t)}{(t-M_Z^2)^{2}} \cr & & 
- \frac{\Pi_{\gamma Z}(t)}{t(t-M_Z^2)} \biggr ]
[\bar u_R \gamma^{\alpha} u_R]
[\bar{\nu}_{\mu}\gamma_{\alpha}(1-\gamma_{5})\nu_{\mu}]
\label{a1loop}
\eeqs

Next, in principle, by measuring the difference $[d\sigma(\nu_\mu +
e_R)/dt+d\sigma(\nu_\mu+e^c_L)/dt]- [d\sigma(\nu_\tau +
e_R)/dt+d\sigma(\nu_\tau+e^c_L)/dt]$, and using the fact that $F^{ee}_Z$,
$\Pi_{ZZ}$, and $\Pi_{\gamma Z}$ are independent of neutrino type, one can
extract the difference of $\lim_{t \to 0} [F_1(t)/t +
F^{\nu\nu}_{Z}(t)/(t-M_Z^2)]$ for $\nu_\mu$ minus $\nu_\tau$ \cite{bpv}; this
is gauge-invariant and is denoted $(1/6)\Delta \langle r^2 \rangle_{EW,\nu_\mu
\nu_\tau}$. However, this does not allow one to isolate $\Delta
F_1^\prime(0)_{\nu_\mu \nu_\tau} =
F_1^\prime(0)_{\nu_\mu}-F_1^\prime(0)_{\nu_\tau}$.  Indeed, explicit
calculation (eqs. (2.30),(2.54) of \cite{ls}; see also \cite{lrz}) shows that
$\Delta F_1^\prime(0)_{\nu_\mu \nu_\tau}$ contains a gauge-dependent term which
diverges as $(g^2/(2^8 \pi^2 M_W^2))[(m_\mu^2-m_\tau^2)/M_W^2] \ln(1/\xi)$ as
$\xi \to 0$.  Hence one cannot isolate a physical $\Delta
F_1^\prime(0)_{\nu_\mu \nu_\tau}$, much less the individual $\nu_\ell$ NCR's.
Note that BPV showed $F_Z^{\nu\nu}(0)=0$ with their pinch technique (to try to
show the observability of $\Delta F_1^\prime(0)_{\nu_\mu \nu_\tau}$) only up to
terms of order $m_\ell^2/M_W^2$, but one cannot neglect the gauge-dependent
$O(m_\ell^2/M_W^2$) terms, which can be made arbitrarily large by gauge choice.
In the SM, $\Delta \langle r^2 \rangle_{EW,\nu_\mu \nu_\tau}$ describes the
interaction with the hypercharge gauge field $B_\alpha$, not the photon; its
gauge invariance allows one to consider the leading term (from eq. (2.57) of
\cite{ls}), $g^2/(16\pi^2M_W^2)[\ln(M_W^2/m_\mu^2) - \ln(M_W^2/m_\tau^2)]$.  We
conclude that, contrary to the BPV claim, one cannot extract a gauge-invariant
NCR.

Since a motivation is to probe for new physics, we note that, in general, using
the reaction $\nu_\mu + e_R \to \nu_\mu + e_R$ does not simplify the analysis.
Consider beyond-SM theories with strong-EW gauge groups $G_{LR} = {\rm SU}(3)_c
\times {\rm SU}(2)_L \times {\rm SU}(2)_R \times {\rm U}(1)_{B-L}$ and
$G_{422}={\rm SU}(4)_{PS} \times {\rm SU}(2)_L \times {\rm SU}(2)_R$
(PS=Pati-Salam).  Here, the first step in the extraction process fails; the
$A_L^\pm$ and $A_R^\pm$ mix to form the mass eigenstates $W_{1,2}^\pm$, and
hence one is not able to remove the 2-$W$ exchange diagrams by considering
$\nu_\mu e_R \to \nu_\mu e_R$.  Neutrino masses affect neutrino electromagnetic
properties (e.g. \cite{ls,fs}), and our comment also applies to the
gauge-dependence of the vector NCR (which vanishes anyway for Majorana
neutrinos) and its axial-vector analogue.


\vspace{-8mm}

\begin{references}

\bibitem{bgl}
W. Bardeen et al., Nucl. Phys. {\bf B46}, 319 (1972).

\bibitem{ls}
B.W. Lee, R. Shrock, Phys. Rev. D {\bf 16}, 1444 (1977).

\bibitem{bpv} 
J. Bernabeu, J. Papavassiliou, J. Vidal, Phys. Rev. Lett. {\bf
89}, 101802 (2002); erratum {\it ibid.} {\bf 89}, 229902(E) (2002). 

\bibitem{bpv2}
J. Bernabeu et al., Phys. Rev. {\bf D62}, 113012 (2000); J. Bernabeu,
J. Papavassiliou, J. Vidal, hep-ph/0210055. 

\bibitem{ssm}
S. Sarantakos, A. Sirlin, W. Marciano, Nucl. Phys. {\bf B217}, 84 (1983);
G. Degrassi et al., Phys. Rev. D{\bf 39}, 287 (1989).

\bibitem{fls}
K. Fujikawa, B. W. Lee, A. I. Sanda, Phys. Rev. D {\bf 6}, 2923 (1972).

\bibitem{lrz}
J.Lucio et al. Phys.Rev.D{\bf 29}, 1539 (1984)(agrees with \cite{ls}).

\bibitem{fs}
M.A.B. Beg et al., Phys Rev. D {\bf 17}, 1395 (1978);
K. Fujikawa, R. Shrock, Phys. Rev. Lett. {\bf 45}, 963 (1980);
B. Kayser, Phys. Rev. D {\bf 26}, 1662 (1982); 
R. Shrock, Nucl. Phys. {\bf B206}, 359 (1982). 

\end{references}
\end{document}